\documentclass[a4paper,prb,onecolumn,floatfix]{revtex4-1}  % for review and submission 
\usepackage{graphicx}
\usepackage{epsfig}
\usepackage{latexsym}
\usepackage{lineno}
\usepackage{color,soul}

\begin{document}

\title{Mechanics of thermally fluctuating membranes.}
\author{J. H. Los, A. Fasolino$^*$, and M. I. Katsnelson}
\affiliation{Radboud University, Institute for Molecules and Materials,
                Heyendaalseweg 135, 6525AJ Nijmegen, The Netherlands}
\date{\today}
\maketitle
%\linenumbers
%{\bf
\section{Abstract}
Besides having unique electronic properties, graphene is claimed to be the 
strongest material in nature\cite{booth2008,lee2008}. In the press release 
of the Nobel committee\cite{nobel} it is claimed that a hammock made of a 
squared meter of one-atom thick graphene could sustain the weight of a 4 kg cat. 
More practically important are many applications of graphene like scaffolds\cite{scaffold} 
and sensors\cite{sensor}  which are crucially dependent on the mechanical strength. 
Meter-sized graphene is even being considered as material for the lightsails in the starshot project 
to reach  the star alpha centaury\cite{starshot}. The predicted exceptional strength of graphene 
is based on its very large Young modulus which is, 
per atomic layer, much larger than that of steel. This reasoning however would apply 
to conventional thin plates\cite{Landauelasticity,Timoscenko} but does not take into 
account the peculiar properties of graphene as a thermally fluctuating crystalline 
membrane\cite{bookMIsha,accountchemreas,nelsonribbons,mirlin-arxiv-Hooks}.
It was shown recently both experimentally\cite{spanishnatphys,mcewuenkirigami,bolotinNat} 
and theoretically\cite{Los1} that thermal fluctuations lead to a dramatic reduction of 
the Young modulus and increase of the bending rigidity for micron-sized graphene samples 
in comparison with atomic scale values. This makes the use of the standard F\"oppl-von 
Karman elasticity (FvK) theory for thin plates\cite{Landauelasticity,Timoscenko}  
not directly applicable to graphene and other single atomic layer membranes. This fact 
is important because the current interpretation of experimental results is based on the 
FvK theory. In particular, we show that the FvK-derived Schwerin equation, routinely used 
to derive the Young modulus from indentation experiments\cite{spanishnatphys} has to be 
essentially modified for graphene at room temperature and for micron sized samples. 
Based on scaling analysis and atomistic simulation we investigate the mechanics of 
graphene under transverse load up to breaking. We  determine the limits of applicability of the 
FvK theory and provide quantitative estimates for the different regimes.
%}

\section{Introduction}
The deflection of a thin plate under transverse point or uniform load 
is normally well described by the FvK
equations. These form a set of two coupled partial differential 
equations reading:
\begin{equation}
\label{FvK1}
\kappa \Delta^2 h - 
\left( 
\frac{\partial^2 {\phi}}{\partial y^2} \frac{\partial^2 h}{\partial x^2} +
\frac{\partial^2 {\phi}}{\partial x^2} \frac{\partial^2 h}{\partial y^2} -
\frac{\partial^2 {\phi}}{\partial x \partial y} \frac{\partial^2 h}{\partial x \partial y}
\right) = P
\end{equation}
\begin{equation}
\label{FvK2}
\Delta^2 \phi - Y \left( 
\frac{\partial^2 h}{\partial x^2} \frac{\partial^2 h}{\partial y^2} -
\left( \frac{\partial^2 h}{\partial x \partial y}  \right)^2 \right) = 0
\end{equation}
where $ h $ is the displacement in the direction perpendicular to 
the plane, i.e. the deflection, $ \phi $ is the potential for the 
in-plane stress tensor,  $ \kappa $ 
is the bending rigidity, $ Y $ the two-dimensional (2D) Young modulus and $ P $ the
transverse pressure. 
%It should be noted that Eqs. \ref{FvK1} and \ref{FvK2} 
%do not include prestress, hence the in-plane stress vanishes for $P=0$. 
%We will first treat this case and then make the extension with prestress 
%later on. 

The behavior of $ h $ as a function of $ P $ and system size $ L $ 
can be obtained from a scaling analysis of the FvK equations as follows. 
Eq. \ref{FvK2} implies that $ \phi/L^4 \sim Y h^2/L^4 $ 
or $ \phi \sim Y h^2 $. Then the second term in eq. \ref{FvK1} 
scales as $ \phi h/L^4 \sim Y h^3/L^4 $ and dominates over the first 
term ($ \sim \kappa h/L^4 $) in the regime of pressures yielding
$ h^2 >> \kappa/Y $. For graphene, with $ \kappa \simeq 1.1 $ eV and 
$ Y \simeq 19.6 $ eV/ \AA$^2$ at room temperature\cite{Fasolino1,Los1},  
%yielding $ \kappa/Y  $ = 0.056 \AA$^2 $, 
this condition implies $ h >> 0.23 $ \AA~ and is normally fulfilled 
for a system of mesoscopic size (or beyond), except for very low pressure.
Hence, apart from this very low pressure regime, the deflection behaves as:
\begin{equation}
\label{hPEq}
h \simeq \left ( \frac{L^4 P}{g Y} \right)^{1/3}
\end{equation}
where $ g $ is a dimensionless number depending on the shape of 
the 2D system and on the type of load, for instance uniform pressure 
or point load with a tip as in nano-indentation~\cite{lee2008}. In the latter
case, the pressure $ P $ in eq. \ref{hPEq} should be replaced by
$ 4 F/(\pi L^2) $ with $ F $ the force exerted by the tip.
Then, eq. \ref{hPEq} turns into an equation which is equivalent to 
the so-called Schwerin equation, but without prestress. The Schwerin
equation is commonly used to determine the elastic modulus $ Y $ from 
nano-indentation measurements\cite{lee2008,spanishnatphys}. From now on 
we will consider a circular drum of radius $ R=L/2 $, clamped at the edge, 
with $ h $ being the midpoint deflection. In that case, the value of $ g $ for 
point load has been derived from analytical solutions of the governing 
equations and is given by $ g = 16/( \pi \tilde{g}_{\nu}^3 )$, with 
$ \tilde{g}_{\nu} \simeq 1.0491-0.1462 \nu - 0.1583 \nu^2 $ 
and $ \nu $ the Poisson ratio\cite{Komaragiri}. With $ \nu \simeq 0.26 $ 
for graphene, one finds $ \tilde{g}_{\nu} \simeq 1.0004 $ and $ g \simeq 5.087 $. 
For uniform load, a fully analytical solution is not available, 
but there are various approximate, semi-analytical solutions. 
The solutions reported in Refs. \cite{Timoscenko,Komaragiri} yield
$ g_{\nu} \simeq 0.7179-0.1706 \nu - 0.1495 \nu^2 \simeq 0.663 $
which is similar to the value obtained in Ref \cite{Yue} yielding
$ g_{\nu} \simeq 75(1- \nu^2) / (8 (23+18 \nu -3 \nu^2) ) \simeq 0.686 $.
It was noted \cite{Yue}, however, that these expressions underestimate 
by 10 \% the values obtained from the classical, more complex and 
accurate solution by Hencky \cite{Hencky} for the case that $ \nu = 0.16 $. 
Here we will use $ g_{\nu} = 0.714 $ yielding $ g \simeq 43.9 $. 
With this larger value for $ g_{\nu} $, compatible with Hencky's solution, 
the simulation data for $ h $ versus $ P $ yield the correct (known) value 
of $ Y $ which agrees with that obtained from a simulation with point load.
We will comment on this later on.
% Without this correction we would have $ g= 54.9 $.

At low  enough pressure, where $ h^2 << \kappa/Y $, we have a linear 
regime where $ h \simeq L^4 P/(f \kappa ) $ according to the above scaling 
analysis with $ f $ another dimensionless, numerical factor. 
From the work in Ref.\cite{Timoscenko,Komaragiri}, identifying the 
bending stiffness $ Y_{3D} d^3/(12(1-\nu^2)) $ for a thin plate 
with bulk Young modulus $ Y_{3D} $ and thickness $d$ as $ \kappa $
for a membrane of atomic thickness, we can evaluate $ f \simeq 1024.0 $ 
and $ f = 256.0 $ for uniform and point load respectively. 

Summing up the two terms from the scaling analysis, one finds 
that $ h(P,L) $ satisfies the equation:
\begin{equation}
\label{CubEq}
f \kappa h + g Y h^3 = L^4 P
\end{equation}
in agreement with the analysis in Refs.\cite{Komaragiri,Wang}.
If we define the cross-over pressure $ P_{c1} $ as the pressure 
for which the linear term equals the non-linear term in eq. \ref{CubEq}, 
which is the case when $ h^2 = f \kappa/(g Y) $, we find that the linear
regime vanishes rapidly with system size as 
$ P_{c1} \simeq 2 \sqrt{(f \kappa)^3/(gY)}/L^4 $.

The above analysis is based on the assumption that the elastic moduli 
$ \kappa $ and $ Y $ are constant, i.e. independent of the system size.
Recently it has been clearly confirmed\cite{Los1}, however, that the 
elastic moduli of graphene are not material constants but scale as
power-laws of the system size due to strong anharmonic coupling between 
in-plane modes and large out-of-plane modes, as predicted by membrane theory
\cite{Nelson1}. Besides, the moduli exhibit an anomalously  strong dependences 
on strain\cite{Los1}. Thus, generally speaking, for a 2D thermally fluctuating solid 
the above analysis is invalid and has to be adapted. Eventually this will lead 
to an anomalous deflection versus load relation $ h \sim P^{\alpha} $ with 
$ \alpha $ different from 1/3 (eq.\ref{hPEq}), as we will show explicitly below.  
Besides analytical results based on a revised scaling analysis for membranes, 
we also present the results from atomistic simulations for a graphene 
drum under uniform load, to validate our analytical findings.

\section{Results: scaling theory}
In order to account for the size and strain dependence of the elastic moduli, 
we extend our scaling analysis by replacing these moduli by their 
renormalized values $ \kappa_R $ and $ Y_R $.
%which depend on both the system size $ L $ and on the strain 
%that is built up in the membrane during deflection and that suppresses 
%anharmonicities at length scales larger than $ L_{\sigma} $.
%The renormalized Young modulus $ Y_R $ as a function of $ L $ and $ \epsilon $ reads:
The latter is given by\cite{Nelson1,nelsonribbons}:
\begin{equation}
\label{YRbyY}
\frac{Y_R}{Y} \sim 
\left\{ \begin{array}{lc}
\vspace*{0.1cm}
\displaystyle
\left( \frac{L}{L_G} \right)^{-\eta_u} & ~~ L_G < L < L_{\sigma} \\
\vspace*{0.1cm}
\displaystyle
\left( \frac{L_{\sigma}}{L_G} \right)^{-\eta_u} & ~~ L_G < L_{\sigma} \leq L
\end{array} \right.
\end{equation}
while $ Y_R/Y = 1 $ for $ L < L_G $ and $ L_{\sigma} < L_G $ with 
$ \eta_u \simeq 0.325 $\cite{Los1}, where $ L_G $ is the so-called Ginzburg 
length beyond which the power-law scaling is applicable. The length $ L_{\sigma} $ 
is the size beyond which anharmonicity is suppressed due to tensile strain and is 
given by\cite{Roldan}:
\begin{equation}
\label{Lsigma}
%L_{\sigma} = \left( \frac{\kappa}{\sigma L_G^{\eta}} \right)^{1/( 2 - \eta )} =
%\left( \frac{\kappa}{ g_{\sigma} \epsilon Y L_G^{\eta}} \right)^{1/( 2 - \eta )}
L_{\sigma} = 
%\left( \frac{ (2 \pi)^2 \kappa}{\sigma L_G^{\eta}} \right)^{ \frac{1}{ 2 - \eta }} =
\left( \frac{ (2 \pi)^2 \kappa}{ 2 B \epsilon L_G^{\eta}} \right)^{ \frac{1}{2 - \eta }} =
\left( \frac{ (2 \pi)^2 \kappa}{ f_{\nu} Y \epsilon L_G^{\eta}} \right)^{ \frac{1}{2 - \eta }}
\end{equation}
where $ \epsilon $ is the average strain and where we used the relation
$ 2B = Y/(1- \nu ) = f_{\nu} Y $ with $ f_{\nu} \equiv 1/(1- \nu) \simeq 1.35 $ (see also Supplementary Information S2).
An equation similar to eq. \ref{YRbyY} applies to $ \kappa_R/\kappa $, 
but with $ - \eta_u $ replaced by $ \eta \simeq 1 - \eta_u/2 \simeq 0.8375 $, implying that $ \kappa_R$
increases with size while $Y_R$ decreases with size.

%Note
%that while the membrane is loaded tensily strain is built up which eventually
%suppresses the anharmonicities.

A theoretical estimate for $ L_G $, 
$ L_{G}^{theor} = 2 \pi \sqrt{16 \pi \kappa^2/(3 Y k_B T)} $\cite{Nelson1}, 
yields $ L_G \sim $ 40 \AA~ at room temperature, but from simulations\cite{Los1}, 
$ L_G $ turned out to be about a factor 2 smaller. Therefore,
in the further analysis we will use $ L_G = c_G L_G^{theor} $, where 
$ c_G \simeq 0.415 $ at 300 K is a correction factor resulting from 
analysis of the simulation data for $ Y_R $ as a function of strain 
reported in Ref.\cite{Los1} (see Supplementary Information S1). 

With renormalized elastic constants, 
eq. \ref{CubEq} still holds, but with $ \kappa $ and $ Y $ replaced 
by $ \kappa_R $ and $ Y_R $. Then, we can again determine the cross-over
pressure $ P_{c1} $ imposing equality of the two terms on the left-hand size. 
For small load where $ L_{ \sigma } > L $, applying the first line of 
eq. \ref{YRbyY}, one finds $ P_{c1} \sim L^{-(6 - \eta )/2} \sim L^{-2.58} $.
For larger loads yielding $ L_G < L_{\sigma} < L $, however, $ P_{c1} $ 
acquires a different size dependence:
\begin{equation}
\label{Pc12}
P_{c1} =
% = 2 f \sqrt{ \frac{\kappa^3}{Y} } \frac{1}{L^{(2-\eta)/2}_G L^{(6+\eta)/2}} 
%= \frac{ \tilde{f}~ \kappa^{ \frac{ 1+ \eta }{2} } }{ Y^{ \frac{ \eta }{4} } }
\frac{ \tilde{g}_1 \kappa^{ \frac{8- 8 \eta }{8 - 4 \eta } } 
Y^{ \frac{ 3 \eta - 2 }{8 - 4 \eta } }
(k_B T)^{ \frac{2 + \eta }{8 - 4 \eta } } }
{L^{ \frac{14 - 9 \eta }{ 4 - 2 \eta } } }
\end{equation}
where $ (14 - 9 \eta )/( 4 - 2 \eta ) \simeq 2.78 $
and $ \tilde{g}_1 \simeq 0.55 f^{ \nu_{1} } g^{ \nu_{2} } $
$ (c_G^2 g_{\sigma} g_{ \epsilon })^{- \nu_{3}} $,
with $ \nu_{1} = (10-7 \eta )/( 8 - 4 \eta ) $, 
$ \nu_{2} = (3 \eta -2)/(8- 4 \eta ) $ and  $ \nu_{3} = (2 + \eta )/( 8 - 4 \eta ) $.
The cross-over in the size dependence of $ P_{c1} $ should occur at a 
system size $ L_{c1} $ at another  critical pressure, $ P_{c2} $ where $ L_{\sigma} $ 
is equal to $ L_{c1} $. Explicit expressions for $ P_{c2} $ as a function of $ L $ 
will be given below. For graphene, it turns out that $ L_{c1} $ would be smaller
than $ L_G $, thus in a regime where renormalization does not apply. 
For $ L > L_G > L_{c1} $, $ P_{c2} < P_{c1} $, implying that 
$ L_{\sigma} < L $ at $ P_{c1} $. Therefore, for graphene 
$ P_{c1} $ is always given by eq. \ref{Pc12}, derived from the 
second line of eq. \ref{YRbyY}, and has no cross-over.
Thus, with renormalized elastic moduli, the regime where the 
first term in eq. \ref{CubEq} is dominant still vanishes for $ L \rightarrow \infty $ 
but more slowly, namely as $ P_{c1} \sim L^{-2.78} $ instead of $ L^{-4} $. 

A third critical pressure $ P_{c3} $ is defined as the pressure for which 
anharmonicity is completely suppressed, i.e. where $ L_{\sigma} \leq L_G $ yielding
$ Y_R = Y $. The important observation to make now is that for pressures 
$ P $ within $ P_{c2} < P < P_{c3} $ or equivalently $ L_G < L_{\sigma} < L $, 
$ h $ as a function of $ P $ obeys
a power-law different from that in eq. \ref{hPEq}, due to the renormalization
of the elastic moduli. Indeed, using eqs. \ref{YRbyY} and \ref{Lsigma}, 
$ Y_R $ depends on the strain $ \epsilon $ as:
\begin{equation}
\label{YR2}
Y_R (\epsilon) 
%\simeq Y \left( \frac{L_{ \sigma} }{ L_G } \right)^{-\eta_u}
\simeq Y \left( \frac{ (2 \pi)^2 \kappa }{ f_{\nu} Y \epsilon L_G^2 } \right)^{- \mu}
%Y \left( \frac{ \kappa }{ \epsilon(h) Y L_G^2 } \right)^{\alpha }
\simeq Y \left( \frac{ 16 \pi \kappa c_G^2 f_{\nu} \epsilon }
{ 3 k_B T } \right)^{\mu}
\end{equation}
with $ \mu \equiv (2- 2 \eta )/(2 - \eta ) \simeq 0.2797 $. 
In a similar way one can derive an expression for $ \kappa_R (\epsilon) $.
Substitution of Eq. \ref{YR2} with $ \epsilon \simeq g_{\epsilon} h^2/L^2 $ 
into Eq. \ref{hPEq} with $ Y$ replaced by $ Y_R $ gives a 
self-consistency equation for $ h $ with solution:
\begin{equation}
\label{hPEqMod}
h \simeq 
\left( \frac{ k_B T }{ \kappa } \right)^{ \frac{\mu}{3+2 \mu} }
\left( \frac{ L^{4+ 2 \mu} P }{ \tilde{g} Y } \right)^{ \frac{1}{3+2 \mu } }
%\left( \frac{ k_B T }{ \kappa } \right)^{ \frac{ 2 - 2 \eta }{ 10 - 7 \eta } }
%\left( \frac{ L^{\mu} P }{ \tilde{g} Y } \right)^{ 1/\mu } 
%h \simeq \frac{ L^{ \frac{12 - 8 \eta }{ 10 - 7 \eta } } 
%(k_B T)^{ \frac{ 2 - 2 \eta }{ 10 - 7 \eta } } } 
%{ \tilde{g} Y^{ \frac{ 2 - \eta }{ 10 - 7 \eta } } 
%\kappa^{ \frac{ 2 - 2 \eta }{ 10 - 7 \eta } } }
%~ P^{ 1/\nu } 
%~ P^{ \frac{ 2 - \eta }{ 10 - 7 \eta } } 
\end{equation}
where $ (3+2 \mu) \simeq 3.56 $ and 
$ \tilde{g} \simeq (16 \pi  c_G^2 f_{\nu} g_{\epsilon}/3)^{\mu} g $.
This equation replaces eq. \ref{hPEq} for the case of a 2D solid
exhibiting renormalization of the elastic moduli according to membrane theory.
One should notice that now the relation between $h$ and $P$ involves, apart 
from $ Y $, also $ k_BT/\kappa $, which is natural as this quantity controls
the strength of anharmonic coupling. Notice that eq. \ref{hPEq} is recovered 
from eq. \ref{hPEqMod} for $ \mu = 0 $ (i.e. for $ \eta_u = 0 $). 

\begin{figure}[htb]
\vspace*{0.00cm}
\includegraphics[width=8.25cm,clip]{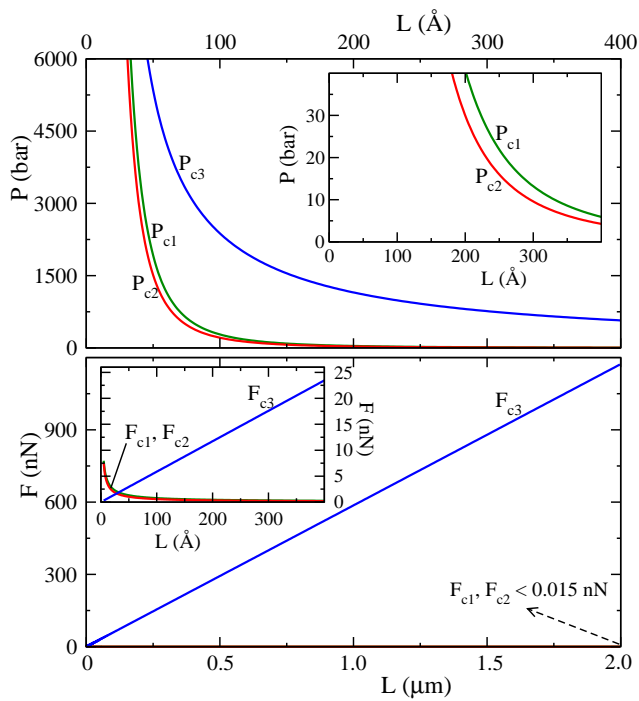}
\caption{\label{PcFc} {\it Calculated critical pressures $ P_{c1} $, $ P_{c2} $, 
$ P_{c3} $ according to eqs. \ref{Pc12}, \ref{Pc2}, \ref{Pc3} for uniform load 
(top panels) and critical forces $ F_{c1} $, $ F_{c2} $, $ F_{c3} $ for point load 
(bottom panels) as a function of system size $ L $ on two different scales, 
in the absence of prestrain. The inset zoom in at different load and/or 
size range.}}
\end{figure}

The geometrical prefactor $ g_{\epsilon} $ for the strain in
$ \epsilon = g_{\epsilon} h^2/L^2 $ depends on the shape of
the deflected membrane. If we define 
$ \epsilon = \sqrt{ A/A_0 } - 1 $ with $ A $ the surface area of the 
deflected membrane and $ A_0 = \pi L^2/4 $ that of the flat drum, then for 
uniform load, with the shape of the drum approximately being that of a spherical cap, 
$ g_{\epsilon} = 2 $, whereas for nano-indentation,
with an approximately cone shaped membrane, $ g_{\epsilon} = 1 $. 

To find the critical pressure $ P_{c2} $, we first 
need to solve $ L_{\sigma} (h) = L $ for $h$. 
Substitution of this $h$ into eq. \ref{CubEq}, retaining both terms 
on the left-hand side with $ \kappa $ and $Y$ replaced by $ \kappa_R $ 
and $ Y_R $ respectively, leads to:
\begin{equation}
\label{Pc2}
P_{c2} = 
\frac{g_{21} (k_B T)^{\frac{3 \eta}{4}} 
\kappa^{\frac{3-3 \eta}{4}} Y^{\frac{3 \eta -2}{4}}}{L^{\frac{8-3 \eta}{2}}} +
\frac{g_{22} (k_B T)^{\frac{7 \eta - 4}{4}} \kappa^{\frac{7-7 \eta}{2}}}
{Y^{ \frac{ 6 - 7 \eta}{4}} L^{\frac{12-7 \eta}{2}}} 
%\frac{g_{21} (k_B T)^{3 \eta/4} \kappa^{(3-3 \eta)/4} Y^{(3 \eta -2)/4}}{L^{(8-3 \eta)/2}} +
%\frac{g_{22} (k_B T)^{(7 \eta -4)/4} \kappa^{(7-7 \eta)/4}}{ Y^{(8-7 \eta)/4}L^{(12-7 \eta)/2}} 
%\frac{ g_2 ~ \kappa^{(7 - 7 \eta)/2}  (k_B T)^{ (7 \eta -4 )/4 } }
%{ Y^{ (6 - 7 \eta )/2 } L^{ (12 - 7 \eta )/2 } }
\end{equation}
with $ ( 8 - 3 \eta )/2 \simeq 2.74 $, $ (12-7 \eta )/2 \simeq 3.07 $,
$ g_{21} \simeq 0.1063 f ( c_G^{3 \eta} f_{\nu} g_{\epsilon} )^{-1/2} $ 
and $ g_{22} \simeq 12.06 g( c_G^{7 \eta - 4} f_{\nu}^3 g_{\epsilon}^3 )^{-1/2} $. 

In a similar way, for deriving $ P_{c3} $ we first solve $ L_{\sigma} (h) = L_G $ 
for $ h $ and then substitute this $ h $ into eq. \ref{CubEq}. 
In this pressure regime the contribution from the term with $ \kappa $ is 
very small (for $L$ not too small) and can be neglected. Then, we obtain:
\begin{equation}
\label{Pc3}
P_{c3} = \frac{g_3 Y}{L} \left( \frac{k_B T}{\kappa } \right)^{3/2}
\end{equation}
with $ g_3 \simeq 0.0146 g ( c_G^2 f_{\nu} g_{\epsilon} )^{-3/2}  $.
%with $ g_3 = g (3/(512 \pi^3))^{3/2} $.
%with $ g_3 \simeq 1.43~10^{-4} $.
%Eq. \ref{Pc3} implies that in the large system size limit, the normal
%$ h(P) \sim P^{1/3} $ behavior is recovered at any pressure, since $ P_{c3} $
%vanishes as $ 1/L $. 
%Eq. \ref{Pc3} reveals a dimensionless material 
%constant $ P_{c3} L/Y = g_3 (k_B T/\kappa)^{3/2} $. 
If we just consider the 
pressure contribution $ \rho mg $ due to the mass of the carbon atoms, with 
$ \rho $ the 2D density of graphene and $ g $ the gravitational acceleration, 
it can directly be calculated from eq. \ref{Pc3} that it requires a system size 
of about $L=305$ km (!) to suppress anharmonicities by graphene's own weight.

The behavior for the various critical loads as a function of system size 
is depicted in Fig. \ref{PcFc} on two different length scales, corresponding 
to the scale used in our simulations and the typical scale in experiments 
respectively. For the latter case we used the parameters for point load 
and displayed critical forces $ F_{ci} = \pi L^2 P_{ci}/4~(i=1,2,3) $
instead of pressures.

\begin{figure}[htb]
\vspace*{-0.25cm}
\includegraphics[width=8.0cm,clip]{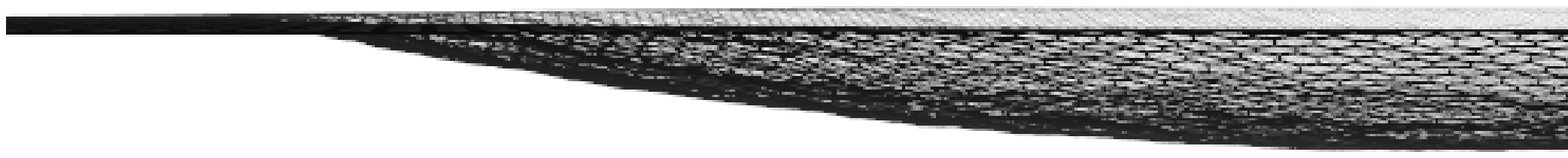}\\
\vspace*{0.2cm}
\includegraphics[width=8.0cm,clip]{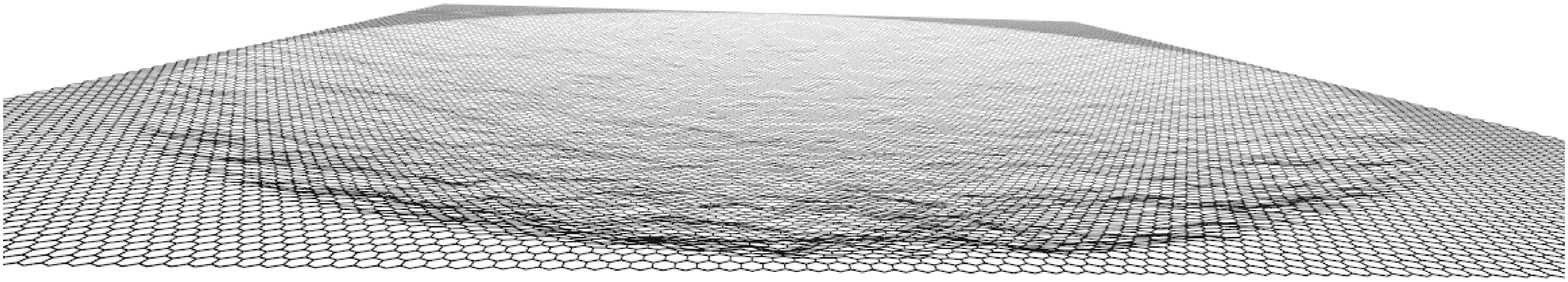}\\
\vspace*{0.2cm}
\includegraphics[width=8.0cm,clip]{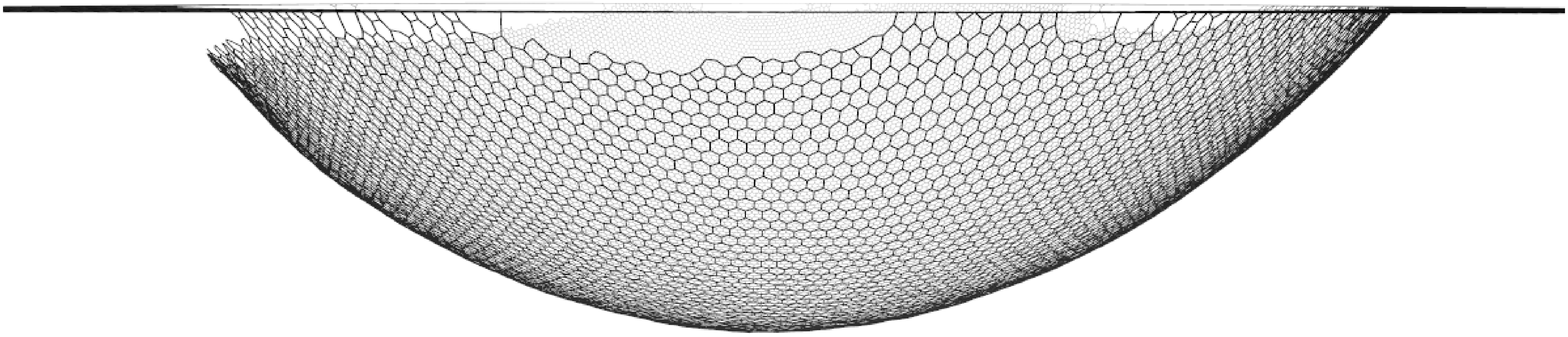}\\
\vspace*{0.2cm}
\includegraphics[width=8.0cm,clip]{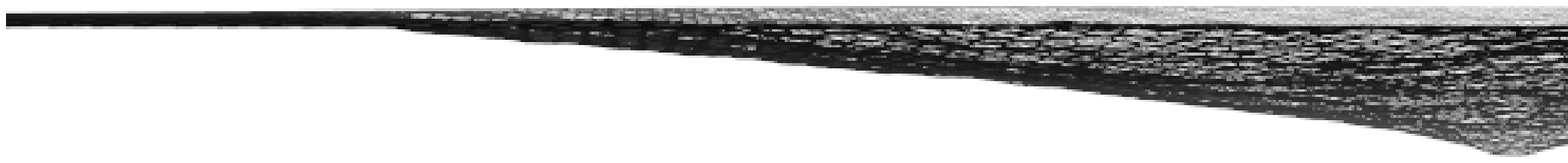}\\
\vspace*{0.2cm}
\includegraphics[width=8.0cm,clip]{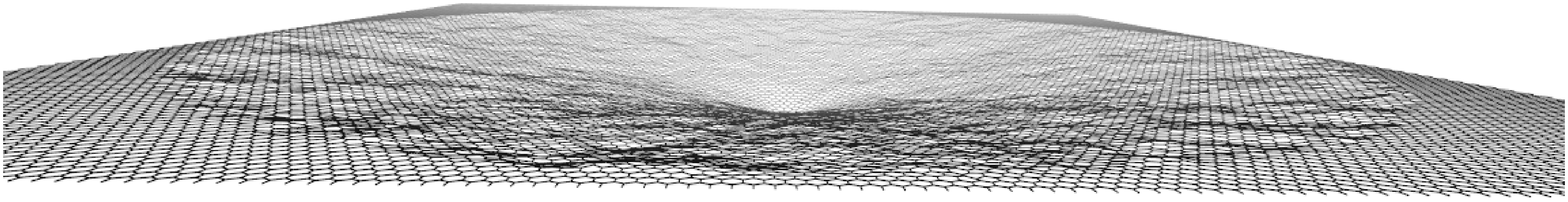}\\
\vspace*{0.2cm}
\includegraphics[width=8.0cm,clip]{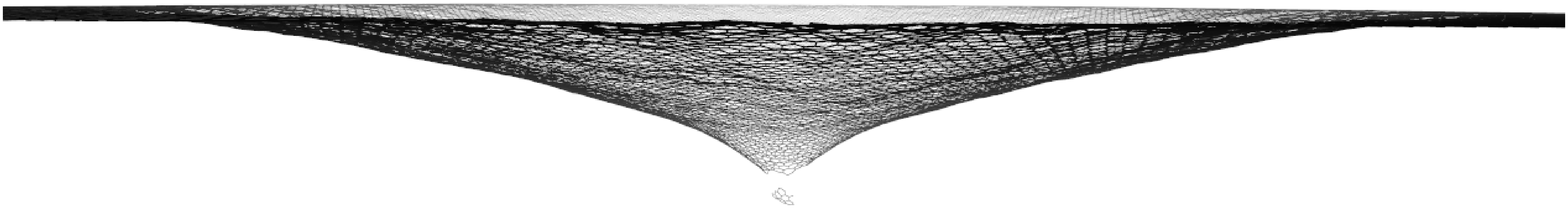}
\caption{\label{Snapshots}
{\it Snapshots from simulations of the indentation of a graphene drum
with diameter $ L \simeq 315 $ \AA~ under uniform load (upper three graphs) 
and point load (lower  three graphs). In the graphs with a view at an angle, 
one can see the thermal corrugation of the drum, responsible for the
anharmonic effects. The bottom graphs show configurations just after
breaking. For uniform load breaking occurs at the drum edge 
at $P \simeq 60 $ kbar while for point load it occurs around the 
tip at $ F \simeq 160 $ nN for the given system size.}}
\end{figure}

\section{Results:atomistic simulations}
In order to validate the behavior derived above and in particular
eqs. \ref{hPEqMod} we have performed atomistic, Monte Carlo simulations 
for a graphene drum with a diameter of $ L \simeq 315 $ \AA~ under uniform
transverse pressures over a wide range between zero and 60 kbar.  
The LCBOPII model was used for the carbon interatomic interactions\cite{Los2}
and the pressure was modelled by assigning a weight $ M = P/(\rho g) $ to each atom. 
Defining the $z-$direction perpendicular to the drum, a change 
$ \delta z $ of the $z$-coordinate of an atom contributes an amount 
$ M g \delta z $ to the energy change $ \Delta E $ of the system entering 
the MC acceptance probability $ P_{acc} = min[1, exp(-\Delta E/k_BT)] $
for the configurational change. The simulations were performed at $ T=300 $ K 
and the 2D density of the drum was adjusted to the equilibrium density
at 300 K, so that no prestrain was present.
For illustration, snapshots from these simulations are shown in Fig. \ref{Snapshots},
together with snapshots from a simulation under point load. In the latter 
case only atoms in a small circular, central region were assigned a weight 
$ M = F/(N_c g) $, with $ F $ the total applied force and $ N_c $ the number 
of atoms in the central circle. For our simulation, $ N_c =25 $.

\begin{figure}[htb]
\vspace*{0.00cm}
\includegraphics[width=8.25cm,clip]{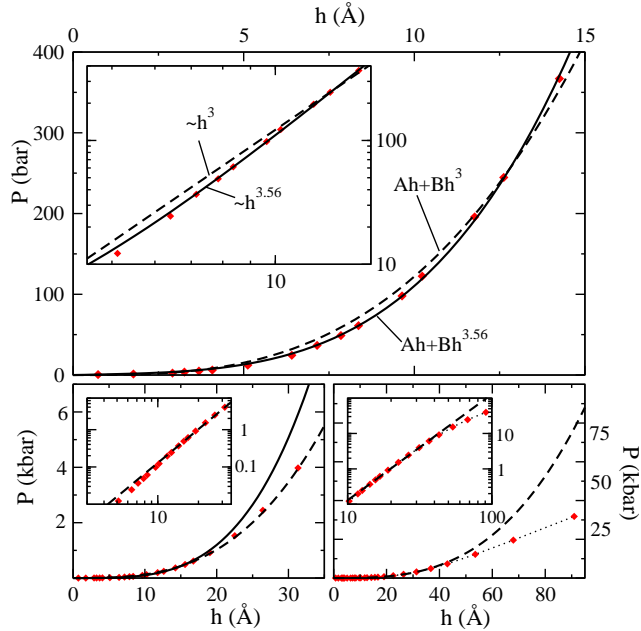}
\caption{\label{Ph_Sim} {\it Pressure as a function of the midpoint 
deflection obtained from simulations (symbols) on three different 
deflection ranges. The solid line is a best fit according to eq. \ref{hPFitMod}
over de interval $ h \in [0,15] $ \AA, while the dashed lines are best
fits according $ P=Ah+Bh^3 $ over the interval $ h \in [0,15] $ \AA 
(upper panel) and $ h \in [15,35] $ \AA (lower panels) respectively.
This insets are the same plots in log-log scale.}}
\end{figure}
The simulation results for uniform pressure $ P $ versus $ h $ are given 
in Fig. \ref{Ph_Sim}. Before analyzing these data we should realize 
that for the derivation of eq. \ref{hPEqMod} we have tacitly neglected the 
linear term in eq. \ref{CubEq}, which is justified for system sizes
commonly used in experiments of the order of 1 $ \mu$m or larger, yielding 
$ P_{c1} < 0.003 $ bar. The system size $ L \simeq 315 $ \AA~ used in our 
simulations, however, requires to include the linear regime to cover the 
pressure range smaller then $ P_{c1} \simeq 12 $ bar in this case. 
While for $ P < P_{c2} $, $ P $ as a function of $ h $ should behave 
as $ P = Ah + Bh^3 $, for $ P > P_{c2} $ the expected behavior is 
$ P = Ah^{(2-3\eta)/(2-\eta)} + Bh^{3+2 \mu } \simeq Ah^{-0.441} + Bh^{3.559} $.
In order to fit our simulation data for pressures below and above 
$ P_{c2} $ we used the following form which combines the usual FvK linear term with the 
renormalized expressions yielding correct asymptotic for $ h \rightarrow 0 $
and $ h \rightarrow \infty $:
\begin{equation}
\label{hPFitMod}
P = A h + B h^{3+2 \mu}
\end{equation}
It has two fitting parameters $ A $ and $ B $ of which the latter is related 
to the elastic moduli by $ B=(\kappa/k_B T)^{\mu} \tilde{g} Y/L^{4+2 \mu } $.

The upper panel in Fig. \ref{Ph_Sim} shows that for pressures up to $\sim$400 bar, 
the simulation data are in good agreement with eq. \ref{hPFitMod} as shown 
by the best fit (solid line), and clearly deviate from a best fit based 
on $ P=Ah+Bh^3 $ expected without renormalization of the elastic constants. 
Instead, beyond 400 bar up to about 6000 bar shown in the 
left bottom panel, the data points are best fitted by $ P=Ah+Bh^3 \simeq Bh^3$, 
which according to our analysis should apply for $ P > P_{c3} $. This
suggests that $ P_{c3} \simeq $ 400 bar, about a factor 2 smaller 
than the estimate from eq. \ref{Pc3} (see also Fig. \ref{PcFc}) but
nevertheless of the right order of magnitude.
The numerical discrepancy here might be due to the fact that cross-over 
regimes are ignored in the theoretical derivations.

The best fit for pressures in the range [400, 40000] bar, beyond $ P_{c3} $, 
based on $ P=Ah+Bh^3 $ yields $ B \simeq 0.128 $, implying a (bare) Young elastic
modulus $ Y = L^4 B/g \simeq 299 $ N/m. This is the value after the mentioned 
adjustment of $ g_{\nu} $ such that it was equal to the $ Y $ obtained
from a point load simulation at an applied force beyond $ F_{c3} $.
The found value is close to the known true bare modulus $ Y=314 $ N/m
at 300 K for LCBOPII. From the best fit at low pressures ($ < P_{c3}$) based 
on eq. \ref{hPFitMod} and with $ \kappa = 1.1 $ eV we obtain $ Y= 307 $ N/m.
The small difference with the value from the first method may be due an uncertainty 
in the factor $ c_g $. In fact, the above two ways to determine $ Y $ (for known 
$ \kappa $) can alternatively be used to determine $ \tilde{g} $ and by this $ c_G $ 
(and $ L_G $) by imposing equality of $ Y $. 

\section{Discussion}
A way the extract the renormalization of $ Y $ from the 
simulation data is to make a fit based on eq. \ref{hPFitMod} and then use that
$ B h^{ 3+2 \mu } = Y_R h^3/(g L^4) $ to obtain the strain dependent $ Y_R $:
\begin{equation}
\label{YRfromFit}
Y_R ( \epsilon ) =  \frac{ L^{4+2 \mu } B }{ g } \left( \frac{h^2}{L^2} \right)^{\mu} =
\frac{ L^{4+2 \mu } B }{ g g_{\epsilon}^{\mu} }~ \epsilon^{\mu} 
\end{equation}
valid for $ \epsilon < \epsilon_{c3} $, where $ \epsilon_{c3} \simeq 0.005 $ is the 
strain required to suppress anharmonicity, i.e. the strain beyond which the normal 
$ P \sim h^3 $ is applicable. Notice that this approach does not require knowledge 
of $ \kappa $ nor $ \tilde{g} $.
%Alternatively, one can deduce $ Y_R $ by calculating $ Y $ from fits based on 
%$P=Ah+Bh^3 $ over a series of pressure domains $ [ 0,P_{max} ] $ as a function of $ P_{max} $.
%The results of these two approaches to recover $ Y_R $ are shown in Fig. \ref{REF}.
For point load, the same approach can be used, but with an additional factor 
$ \pi /(4 L^2) $ multiplying the right-hand side of eq. \ref{YRfromFit}.

%for pressures below $ P_{c3} $ yields an only slightly 
%different value $ Y \simeq 218 $ N/m.

The right bottom panel shows a deviation from the $ P \sim h^3 $ 
behavior for deflections beyond 40 \AA. 
This deviation for large strain can be attributed to a normal softening of the elastic moduli 
due to stretching anharmonicity, as in 3D crystals. Such a softening for graphene 
under large strain is in agreement with previous observations \cite{Gao}.
The corresponding critical pressure, $ P_{c4} $, should 
depend on the size as $ P_{c4} = g_4/L $, with $ g_4 \simeq 25.2 $ N/m derived 
from the value $ P_{c4} \simeq 8 $ kbar ($8~10^8$ N) for $ L=3.15~10^{-8} $ m.
Indeed, assuming that this anharmonicity sets in at a fixed (size independent) 
critical strain, $ \epsilon_{c4} = g_{\epsilon} h_{c4}^2/L^2 $, the size dependence 
of $ P_{c4} $ follows directly from eq. 4, neglecting the linear term, yielding  
$ P_{c4} \simeq (g Y / L) (\epsilon_{c4}/g_{\epsilon})^{3/2} $, with
$ \epsilon_{c4} \simeq 0.032 $. Although $ P_{c4} $ is a safe lower bound 
for breaking, normally breaking is only expected at significantly 
higher strains, where the in-plane moduli start to vanish due to the anharmonicity
of the interaction potential. For LCBOPII, the bulk modulus vanishes
at a strain value of $\sim$0.2, a value indeed close to the strain where breaking
was actually observed in our simulation, at a pressure $ P_{br} \simeq 50 $ kbar.
This value of the breaking strain is similar to that found in a simulation 
study of graphene nanoribbons under uniaxial strain \cite{Lu}.
It should be noticed, however, that a typical atomistic simulation only covers 
a very small time interval, typically orders of magnitude smaller than a second, 
which makes the choice of a maximal, safe lower bound for breaking from simulations
at a given temperature not obvious and somewhat arbitrary. 

Staying on the safe side by choosing the breaking pressure as $ P_{br} = 4 P_{c4} = 4 g_4/L $, 
yielding $ P_{br} \simeq $ 32 kbar for the simulated system size with a corresponding 
strain of $\sim$0.13, a graphene drum of 1 m in diameter gives a breaking force of  
$ F_{c4} = \pi L^2 P_{c4}/4 \simeq 79.2 $ N, enough for an extremely heavy cat 
of about 8.0 kg to be safe, treating it as a uniform load. Treating the cat as
a point load, however, and assuming that the breaking strain is equal
to that for uniform load, we have to correct $ g_4 $
by a factor $\sim$ 0.328 due to the different values for $ g $ and 
$ g_{\epsilon} $, implying that the cat should not be heavier than 2.65 kg,
i.e. a young cat, to be safe. 
In reality, a cat on a drum of this size is something between point 
and uniform load, so that probably any cat should be safe on it.
It is interesting to notice, and somewhat counterintuitive, that while
a drum of 1 m cannot bear a person of 100 kg, a drum
of 40 m could, due to the fact that $ F_{c4} $ grows linearly 
with $ L $. A graphene drum would only break by its own weight 
for a size $ L=4 g_4/(\rho mg) \simeq 13520 $ km !

While the relations derived above are appropriate for the analysis 
of our simulations where prestress can be controlled and taken to be zero,
it should be noticed that in nano-indentation experiments almost unavoidably 
some prestress $ \sigma_0 $ is present, created during preparation of the drum. 
The implications for the load versus deflection expression and the various critical
loads for the case of tensile prestress, including renormalization 
of the elastic moduli, are given in the Supplementary Information S3.
Tensile prestress gives rise to a contribution to the force which is linear in $h$, 
namely $ \pi \sigma_0 h $. As this is an order $ L^2 $ larger then the linear 
contribution $ ~ \kappa h/L^2 $ arising from the FvK equations, its contribution 
increases with system size as $ L^{2 \eta -1} $, and can be significant 
as compared to the cubic term $ ~ Y h^3/L^2 $, even for 
$ \mu $m sized drums. Therefore, for the sake of accuracy in measuring 
the elastic modulus and the effect of its renormalization, one should keep the 
prestress as small as possible, so that the term cubic in $ h $, from which the 
elastic modulus is determined, is the dominant term. Moreover, while $ F_{c1} $ 
increases with system size for tensile prestress, $ F_{c2} $ and $ F_{c3} $ decrease 
with $ \sigma_0 $. To be able to measure the renormalization of the elastic modulus, 
however, $ F_{c3} $ should not be too small, leaving a sufficiently large force 
domain ($ < F_{c3} $) for observing renormalization.

\section{Conclusions}
The revised FvK theory for thermally excited membranes like graphene that we have presented here 
is important for any technological application of 2D materials involving their mechanical properties. 
For graphene, we have discussed in a quantitative way the behaviour under uniform and point load up to breaking.

\begin{acknowledgments}
This project has received funding from the European Union’s Horizon 2020 research and innovation programme under grant agreement No. 696656 – GrapheneCore1. We thank Cristina  Gomez-Navarro, Julio Gomez-Herrero and 
Guillermo L/'opez-Pol/'in for interesting discussion.
\end{acknowledgments}


\begin{thebibliography}{99}
\bibitem{booth2008} Booth, T. J., Blake, P., Nair, R. R., Jiang, D., Hill, E. W., Bangert, U., Bleloch, A.,
Gass, M., Novoselov, K. S., Katsnelson, M. I. \& Geim, A. K. Macroscopic Graphene Membranes and 
Their Extraordinary Stiffness. Nano Lett. {\bf 8}, 2442-2446 (2008)
\bibitem{lee2008} Lee, C., Wei, X., Kysar, J. W. \& Hone, J. Measurement of the Elastic Properties and 
Intrinsic Strength of Monolayer Graphene.  Science {\bf 321}, 385-388 (2008)
\bibitem{nobel}
https://www.nobelprize.org/nobel\_\\
prizes/physics/laureates/2010/advanced-physicsprize2010.pdf
\bibitem{scaffold} Vickery, J.L., Patil, A. J. \& Mann S. Fabrication of Graphene–Polymer Nanocomposites
With Higher-Order Three-Dimensional Architectures. Adv. Mater. {\bf 21} 2180–2184 (2009)
\bibitem{sensor} Dolleman, R.J., Davidovikj, D., Cartamil-Bueno, S.J.,   van  der  Zant, H.S.J.,
 \& Steeneken P.G. Graphene Squeeze-Film Pressure Sensors NanoLett.{\bf 16}, 568 (2016)
\bibitem{starshot} https://breakthroughinitiatives.org/Initiative/3
\bibitem{Landauelasticity} Landau L. D. \&  Lifshitz, E. M. {\it Theory of Elasticity},
Pergamon, Oxford 1970.
\bibitem{Timoscenko} Timoshenko, S.P. \& Woinowsky-Krieger,S. 
{\it Theory of Plates and Shells}, (New York: McGraw-Hill, 1951).
\bibitem{bookMIsha} Katsnelson, M.I. {\it Graphene: Carbon in Two Dimensions} (Cambridge Univ. Press, Cambrigde, 2012)
\bibitem{accountchemreas}Katsnelson, M. I. \& Fasolino, A., Graphene as a Prototype Crystalline Membrane
Acc. Chem. Res., {\bf 46}, 97-105 (2013)
\bibitem{nelsonribbons} Kosmrlj A. \& Nelson, D.R, Response of thermalized ribbons to pulling and bending,
Phys. Rev. B {\bf 93}, 125431 (2016)
\bibitem{mirlin-arxiv-Hooks}Gornyi, I. V., Kachorovskii, V. Yu. \&  Mirlin, A. D., Anomalous Hooke's law 
in disordered graphene. arXiv:1603.00398
\bibitem{spanishnatphys}L\'opez-Pol\'in, G., et al., Increasing the elastic modulus of graphene 
by controlled defect creation, Nature Physics {bf 11},26-31 (2015).
\bibitem{mcewuenkirigami}Blees, M. K. et al., Graphene kirigami, Nature {\bf 524}, 204-207 (2015)
\bibitem{bolotinNat} Ryan J.T. Nicholl , Hiram J. Conley , Nickolay V. Lavrik, Ivan Vlassiouk, 
Yevgeniy S. Puzyrev, Vijayashree Parsi Sreenivas, Sokrates T. Pantelides, and Kirill I. Bolotin, 
Mechanics of Free-Standing Graphene: Stretching a Crumpled Membrane, Nature Comm. {\bf 6} 8789 (2015)  
\bibitem{Los1} Los, J. H., Fasolino A. \&  Katsnelson, M. I., Scaling behavior and strain dependence 
of in-plane elastic properties of graphene. Phys. Rev. Lett. {\bf 116}, 015901 (2016)
\bibitem{Fasolino1} Fasolino, A., Los, J. H. \& Katsnelson, M. I., Intrinsic ripples in graphene,  
Nat. Mater. {\bf 6}, 858-861 (2007).
\bibitem{Komaragiri} Komaragiri, U. \& Begley, M. R.,
The Mechanical Response of Freestanding Circular Elastic Films Under Point and Pressure Loads
J. Appl. Mech. {\bf 72}, 203 (2005).
\bibitem{Yue} Yue K., Gao W., Huang R. \& Liechti K. M.,
J. of Appl. Phys. {\bf 112}, 083512 (2012).
\bibitem{Hencky} Hencky H., Z. Math. Phys. {\bf 63}, 311 (1915).
\bibitem{Wang} Wang P., Gao W., Zhiyi Cao Z., Liechti K.M. \& Huang R.,
J. Applied Mechanics {\bf 80}, 040905 (2013).
\bibitem{Nelson1} Nelson, D. R., Piran,  T., \& Weinberg, S. (eds) 
{\it Statistical Mechanics of Membranes and Surfaces}, World
Scientific, Singapore, 2004.
\bibitem{Roldan} Rold\'an, R., Fasolino, A., Zakharchenko, K. V. \& Katsnelson, M. I.,
Suppression of anharmonicities in crystalline membranes by external strain,
Phys. Rev. B {\bf 83}, 174104 (2011).
\bibitem{Los2} Los, J. H., Ghiringhelli, L. M., Meijer,  E. J., \& Fasolino, A.,
Improved long-range reactive bond-order potential for carbon. I. Construction,
Phys. Rev. B {\bf 72}, 214102 (2005).
\bibitem{Gao} Gao W., Huang R., J. Mech. Phys. Solids {\bf 66}, 42 (2014).
\bibitem{Lu} Lu Q., Gao W., Huang R., Modelling Simul. Mater. Sci. Eng {\bf 19}, 
054006 (2011).
\end{thebibliography}
\end{document}